\documentclass[prb,twocolumn,showpacs,floatfix]{revtex4}
\usepackage{graphicx}
\usepackage{dcolumn}% Align table columns on decimal point
\usepackage{bm}% bold math
\begin{document}

\title{Hysteresis loops of magnetic thin films with perpendicular anisotropy}
\author{E. A. Jagla}
\affiliation{The Abdus Salam International Centre for Theoretical Physics\\
Strada Costiera 11, (34014) Trieste, Italy}
\date{\today}

\begin{abstract}
We model the magnetization of
quasi two-dimensional systems with easy  perpendicular ($z$-)axis anisotropy upon 
change of external magnetic field along $z$. 
The model is derived from the Landau-Lifshitz-Gilbert equation for
magnetization evolution, written in closed form in terms of the $z$ component of the magnetization only.
The model includes--in addition to the external field--magnetic exchange, dipolar interactions
and structural disorder.
The phase diagram in the disorder/interaction strength plane is presented, and the different
qualitative regimes are analyzed. The results compare very well with observed experimental
hysteresis loops and spatial magnetization patterns, as for instance for the case of Co-Pt multilayers.

\end{abstract}

\maketitle

\section {Introduction}

Hysteresis is widespread in nature. Its origin and even what we mean by it
may depend on the particular context.\cite{bertotti}
Our concern here is that type of hysteresis appearing 
due to the complex structure of the energy landscape 
of the system studied, when some external control 
parameter is changed. 
In the paradigmatic description of the configuration space, the system is represented by
a point evolving in a complicated energy landscape.
If the available thermal energy is
small compared to the heights of the typical barrier between local minima, the system
can only evolve by moving down in energy and it will remain in a local energy minimum. 
If some external control parameter is changed, the system
adapts to the external conditions in order to reduce its energy. 
Upon changes of the external parameter it moves smoothly following the shift 
of the minimum in which it is 
located, until this minimum eventually becomes unstable. At this point the system
rolls down in energy and finds another minimum in which it stays while this 
is stable. The evolution of the system in passing from one minimum that 
becomes unstable to another minimum implies energy dissipation. If the external field 
performs a cyclic variation, returning to its original value, the energy dissipated
during the process can be calculated as the work done by the external field. If 
the external field $B$ is conjugated to an internal variable $M$ in the system, 
such that there is a term in the (differential) free energy of the system of the form $MdB$, 
then the evolution of the system 
in the $M$-$B$ plane is hysteretic. Magnetic hysteresis is the prototypic example 
of this behavior, in which $M$ is the system magnetization, and $B$ is the external
magnetic field. 
In this description of hysteresis the external field is assumed to change 
infinitely slowly, in such a way that
the system adapts in every moment to the actual value of 
the external field. However, this in general is not enough to guarantee a smooth evolution
of the microscopic constituents of the system, 
and that is the origin of dissipation and at the end, of hysteresis.

Although the qualitative origin of hysteresis is well understood, a completely different issue
is to reproduce in detail the observed hysteretic behavior of particular systems.
For magnetic materials this is an extremely important issue, since hysteresis 
is at the base of magnetic information storage technology, and then its detailed
understanding has extreme 
practical importance. From another perspective, the different macroscopic 
characteristics of the hysteresis loops correspond to different
microscopic evolution of the spatial magnetization distribution. The details of this correlation
are worth to be studied to gain insight into the fundamental physical principles involved.

In the present paper we model the hysteretic behavior of thin magnetic films with perpendicular
anisotropy. In the next Section we discuss how to obtain a closed equation for the evolution
of the perpendicular component of the magnetization, starting from the vectorial 
Landau-Lifshitz-Gilbert equation. The equation we obtain is qualitatively similar to some
phenomenological proposals, but the derivation given here indicates clearly its possibilities
and limitations. Section III contains the results of the numerical simulation, and in 
Section IV we compare the results with experimental data in Co-Pt multilayers and 
conclude.

\section{Derivation of the model}

Magnetism originates in quantum effects occurring at the atomic level. To model 
macroscopic hysteresis however, it is impossible in practice to start from these elemental building blocks
in order to arrive to the macroscopic description. The {\em micromagnetics approach}\cite{bertotti,huber} starts
from an intermediate scale description, in which the magnetic moment of an elemental piece of material
$\bf m$ is taken as the fundamental variable. $\bf m$ is a vector field, which is a function of the 
spatial coordinates and time, and it is assumed to satisfy at any position and time the constrain
$|{\bf m}|^2=1$, where a rescaling of the amplitude has been assumed.
A reasonably detailed description of magnetization evolution is provided by the 
phenomenological Landau-Lifshitz-Gilbert equation,\cite{bertotti,huber,llg} that can be written as
\begin{equation}
\frac{\partial {\bf m}}{\partial t}=-a {\bf m}\times {\bf B}-b {\bf m}\times
\left ({\bf m}\times {\bf B}\right)
\label{llg}
\end{equation}
where ${\bf B}$ is the local effective magnetic field.
%, incorporating external contributions and internal ones that appear
%as a consequence of the interaction in the system. In general ${\bf H}$ is a function of position, time, 
%and the actual distribution of the magnetization in the system at the given time. 
The first term on the right of Eq. (\ref{llg}) describes a precessional evolution
of $\bf m$ around $\bf B$, and the second term is a phenomenological damping term that tends
to align $\bf m$ with $\bf B$. Note that from this equation we obtain immediately that 
$\partial |{\bf m}|^2/\partial t=0$, i.e, it automatically maintains the norm of $\bf m$ as fixed. 
In equilibrium ($\partial {\bf m}/\partial t=0$), this equation reduces to 
Brown's equation,\cite{bertotti,huber} namely  ${\bf m}\times {\bf B}=0$, stating that at any point the local 
magnetization has to be aligned with the local field.
The local field ${\bf B}$ includes contributions from the externally applied field,
from the interaction with all the rest of the magnetic moments in the sample,
and from structural anisotropies in the sample.

Equation (\ref{llg}) is a sufficiently detailed starting point to obtain the time and space evolution of the
magnetization. Unfortunately, it is frequently a too much detailed
starting point. In fact, except in those simple cases in which an analytical approach is possible, the
numerical solution of Eq. (\ref{llg}) is difficult because the orientation of the local field changes in general
with time.

For the problem we want to study we can obtain from Eq. (\ref{llg}) 
a numerically tractable simplified formulation.
The problem is that of the magnetization evolution of (quasi) two-dimensional systems
with perpendicular ($z$) easy axis anisotropy, in the presence of an external magnetic field along $z$.
This problem has attracted a lot of attention, in particular because it constitutes a promising configuration
for high density magnetic recording.\cite{iwasaki}
The perpendicular easy axis anisotropy induces that in most of the sample the magnetization points in the
plus or minus $z$ direction, defining magnetization domains. Only within the `domain walls' separating 
magnetic domains with opposite magnetization, the local magnetization cannot be considered 
to point along $z$. 
In the assumption that the extent of the system occupied by domain walls is small compared to the 
domains themselves, we can derive an effective model from Eq. (\ref{llg}) taking into account only the 
$z$ component of the magnetization.

To derive this model, we first of all need to specify in more detail what the contributions to the local field
${\bf B}$ are. We will consider the following four contributions to ${\bf B}$:
\begin{equation}
{\bf B}=h_0 \hat {\bf z} +Am_z \hat {\bf z} -\gamma \int d{\bf r'} \frac{{\bf m}({\bf r'})}{|r-r'|^3}
%+J_0\sum_j{\bf m}({\bf r}_j)
+{\bf B}_{exch}
\label{h}
\end{equation}
The first term is an external field applied along $z$. Second term is the lowest order contribution
from the assumed easy axis anisotropy of the sample. Third term is the dipolar (or stray) field
generated by all the rest of the sample. The last term accounts for the exchange field.
It typically consists of a sum of the magnetization values on neighbor sites, 
and then it introduces an explicit dependence on the symmetry of the
underlying lattice. In the simplest case in which we want to get rid of lattice effects
to concentrate on the intrinsic dynamics of the model, it is convenient
to have a symmetric form for the exchange term.
By noticing that if we add to $\bf B$ a term proportional to the local magnetization
it does not
affect the equation of motion (since only ${\bf B}\times {\bf m}$ 
appears), the exchange term can be written in the form
\begin{equation}
{\bf B}_{exch}=J\nabla ^2 {\bf m}
\end{equation}
where $J$ defines the exchange constant.
Note that the fact that we take $|{\bf m}|=1$ is incorporated in the values of $A$, $\gamma$ and $J$.

To obtain a closed equation for the evolution of $m_z$, we notice that under the assumptions made, the
dipolar field points to a good approximation along the $z$ direction everywhere
(since non-$z$ components are originated
in the field generated by the domain walls, which is much smaller).
Then the first three terms of Eq. (\ref{h}) point in the $z$ direction. We separate explicitly 
this part from the exchange part:
\begin{eqnarray}
{\bf B}&\simeq&\left (h_0  +Am_z -\gamma\int d{\bf r'} \frac{ m_z({\bf r'})}{|r-r'|^3}\right )\hat {\bf z}
+J\nabla^2 {\bf m}\nonumber\\ 
&\equiv& h\hat {\bf z} +{\bf B}_{exch}
\label{hh}
\end{eqnarray}
from these two terms we get two contributions to the $z$ component of Eq. (\ref{llg}):

\begin{equation}
\frac{\partial {m_z}}{\partial t}=b (1-m_z^2)h -
\left . a {\bf m}\times {\bf B}_{exch}\right |_z -b \left .{\bf m}\times
\left ({\bf m}\times {\bf B}_{exch}\right)\right |_z
\label{hz}
\end{equation}
Although within the domains ${\bf B}_{exch}$ points in the $z$ direction, it certainly does not
within domain walls. The ${\bf B}_{exch}$ term will influence the detailed structure of the domain
walls. Since we are not interested in the internal structure of the walls, we can proceed
by replacing ${\bf B}_{exch}$ by some phenomenological expression dependent on $m_z$ only.
One way of obtaining such an expression is the following. Consider a domain wall extending along the $y$ axis. 
If we imagine that the internal structure of the wall is that known as a Bloch's wall, 
then $m_x=0$. On the contrary, if we think that the wall is a Neel's wall we
have $m_y=0$.\cite{huber}
In both cases magnetization can be described in terms of the angle $\theta$ between local magnetization 
and the $z$ axis, in
such a way that $m_z=\cos(\theta)$, and $m_{y}=\pm \sin(\theta)$, $m_x=0$ for a Bloch's wall, or
$m_{x}=\pm \sin(\theta)$, $m_y=0$ for a Neel's wall. In any of these cases the second term in
the r.h.s. of Eq. (\ref{hz}) still vanishes, and the last term takes the 
simple form $bJ\sin (\theta) \frac{d^2\theta}{dx^2}$, 
that can be re-expressed in terms of $m_z$ and reinserted into (\ref{hz}) to give

\begin{equation}
\frac{\partial {m_z}}{\partial t}=b \left[ (1-m_z^2)h +
J\frac{d^2m_z}{dx^2}+J\frac{m_z}{1-m_z^2}\left (\frac{dm_z}{dx}\right)^2\right ]
\end{equation}
Once in this form, the equation can be generalized to describe a system with 
domain walls oriented in any direction, in the form
\begin{eqnarray}
\frac{\partial {m_z}}{\partial t}=b\left [ (1-m_z^2)\left (h_0  +Am_z -\gamma\int d{\bf r'} \frac{ m_z({\bf
r'})}{|r-r'|^3}\right ) \right .+\nonumber\\
\left .+J\nabla^2m_z+J\frac{m_z}{1-m_z^2}(\nabla m_z . \nabla m_z)\right ]
\label{completa}
\end{eqnarray}
where we explicitly introduced the form of $h$ from Eq. (\ref{hh}).
As we explained above, this equation will be meaningful only in 
cases in which the magnetization distribution consists of domains of positive an 
negative magnetization, separated by domain walls, and such that
the thickness of the domain walls is small compared to the typical width of the domains. 
We first present some simple examples of the use of this equation, and then the results of numerical simulations
to study the form of hysteresis loops of  two dimensional garnets.

The first simple example is that of a single, isolated spin in the presence of an external field ${\bf h}_0$
which is fixed in orientation (along $z$), although its amplitude may depend on time ${\bf h}_0=h_0(t) \hat {\bf z}$. 
The only term in the r.h.s. of Eq. (\ref{completa}) that survives in this case is the first one, 
and the evolution equation is:
\begin{equation}
\frac{\partial m_z}{\partial t}=b(1-m_z^2)h_0(t)
\end{equation}
that can be immediately integrated to give

\begin{equation}
m_z(t)=\tanh \left (b\int h_0(t)dt\right)
\end{equation}
The particular case in which $h_0$ is constant describes the spin flip process of an isolated spin in a constant
field.

The second example is the stationary structure of a straight domain wall in the presence of structural
anisotropy, with no dipolar or external field. The equation to be solved becomes
\begin{equation}
A(1-m_z^2)m_z=J\left [\frac{d^2m_z}{dx^2}+\frac{m_z}{1-m_z^2}\left (\frac{dm_z}{dx}\right)^2\right]
\label{a}
\end{equation}
which in terms of the variable $\theta$ (such that $\cos(\theta)=m_z$) is written as
\begin{equation}
\frac{d^2\theta}{dx^2}=\frac A J \sin(\theta)\cos(\theta)
\end{equation}
and from here the well known solution
\begin{equation}
m_z=\cos(\theta)=\tanh \left ( x\frac AJ \right )
\end{equation}
is obtained.
We see from here how the thickness of the wall is proportional to the ratio $J/A$ between the 
exchange and anisotropy constants.
It can be verified that the same spatial dependence but with a domain wall thickness which is
doubled, is a solution of  Eq. (\ref{a}) if we drop the last term, quadratic in $dm_z/dx$. 
Below we will return to this point.

Before moving to the simulations of two dimensional systems,  we will compare the present model with 
others that have been proposed to describe
similar problems. Starting from the basic Ising model of discrete $\pm 1$ spins, Sethna {\em et al.} \cite{sethna}
have introduced the random field Ising model (RFIM). This is an important model system, 
where the effect of disorder on
the existence or not of an abrupt nucleation step was analyzed in detail. Other people have considered the
effect of adding dipolar interaction to the RFIM. The dipolar RFIM \cite{magni} contains in principle all
physical ingredients necessary to provide a good description of uniaxial anisotropy systems (not necessary
thin films) in the presence of magnetic field along the easy direction. However, it suffers from a technical
problem, namely that since the fundamental variable is discrete, domain walls are forced to be one lattice
parameter thick. This produces domain walls that are artificially pinned to the numerical lattice, and 
this makes realistic simulations very difficult. An alternative to overcome this problem consist in 
using a continuum variable $\phi$, which instead of taking only the values $\pm 1$, can have intermediate values,
at the cost of paying some energy.
We obtain in this way what can be called the ``dipolar phi-fourth" model, described by the equation\cite{seul,sagui,jagla}

\begin{equation}
\frac{\partial\phi ({\bf r})}{\partial t}=
h_0-A (-\phi+\phi^3) 
-\gamma\int d{\bf r}' \frac{\phi({\bf r}') }{(|{\bf r}-{\bf r}'|^3)}+J \nabla^2 \phi 
\label{fi4}
\end{equation}
The scalar variable $\phi$ (which may be assimilated to the $z$ component of the local magnetization) 
evolves in a local two-well potential which mimics the existence of two privileged values ($\pm 1$)
for $\phi$. External field, dipolar and exchange contributions are included.
We see a general similarity between this equation and our Equation (\ref{completa}). 
However, a few differences need to be 
mentioned. The most obvious one is that in Eq. (\ref{fi4}) $\phi$ is not restricted to satisfy
$|\phi|<1$ and in fact, larger values are obtained, in particular if the external field is strong enough.
In Eq. (\ref{completa}), the restriction $|m_z|<1$ is guaranteed by the $(1-m_z^2)$ factor.
A second difference is in the structure of the
gradient terms: the last, non-linear term in Eq. (\ref{completa}) 
does not appear in Eq. (\ref{fi4}).
Although in some cases this term may have some 
physical importance, we have seen that for a simple, static domain wall,
it only renormalizes its thickness.
In view of the difficulties
that the presence of this term would add to the simulation (making them much more time consuming because of
its non-linear nature)
we have decided to drop it in the simulations that are presented below, 
and assume that its main effect has
been incorporated in an appropriate renormalization 
of the amplitude of the $\nabla ^2 m_z$ term. This simplifying assumption is justified also since we do
not intend to model details of the interior of the domain walls, and then all that we need is a term
that produces some energy cost for the domain wall. The $\nabla ^2 m_z$ term suffices in this
respect.

\section{Results}

The model simulated is thus given by the equation
\begin{equation}
\frac{\partial {m_z}}{\partial t}= (1-m_z^2)\left (h_0  +A({\bf r})m_z -\gamma\int d{\bf r'} \frac{ m_z({\bf
r'})}{|r-r'|^3}\right ) +J\nabla^2m_z 
\label{completa2}
\end{equation}
which is Eq. (\ref{completa}) in the absence of the last term, 
where time has been rescaled to absorb the constant $b$,
and where a spatial dependence of the coefficient $A$ has been made 
explicit. This dependence will allow us to simulate structural disorder in the sample. 
A single, isolated spin, is in equilibrium if it points in the
direction of the local field.
We see that $m_z=+ 1$ ($m_z=- 1$) is an equilibrium configuration of the spin 
if $h_0+A>0$ ($h_0-A<0$). Then $A$ represents the value of the local coercive field that is necessary to apply in order 
to invert the orientation of the spin. The function $A({\bf r})$ will be chosen
in the form 
\begin{equation}
A({\bf r})=A_0(1+D\eta({\bf r}))
\label{desorden}
\end{equation}
where $\eta$ is a spatially
random and uncorrelated Gaussian variable, with zero mean and unitary variance ($\eta$ is cut off
at large negative values, namely $\eta>-1/D$, to guarantee $A>0$).
$D$ controls the overall intensity of disorder. The parameters of the model are thus
$A_0$, $D$, $\gamma$ and $J$, in addition to the external field $h_0$.
As explained in [9], %\cite{jagla}, 
the ratio between $J$ and $\gamma$ can be adjusted 
by appropriately rescaling the spatial scale in the model.
In the simulations below we choose $\gamma=0.095~J$, and give the results in term of $h_0/A_0$, $J/A_0$, and
$D$. The mesh parameter is taken as the unit of length.

The results presented below correspond to zero temperature. Thus the evolution of the system is driven
exclusively by the tendency to minimize the energy, as described in the introductory part. We will see that 
even this $T=0$ case is very rich, and provides different macroscopic forms of the hysteresis curve when the
interaction strength between spins and the amount of structural disorder are changed. The 
results obtained compare very
well with experimental results.

\begin{figure}
\includegraphics[width=8.5cm,clip=true]{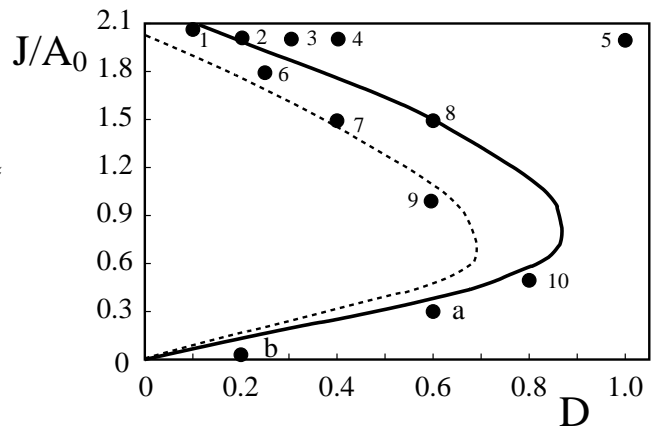}
\caption{\label{phasediag}
Disorder-exchange/anisotropy ratio ($D-J/A_0$) phase diagram for the model described by
Eq. (\ref{completa2}), in a system of size $384\times384$, with $\gamma=0.095~J$. Numbers indicate 
the points for which the hysteresis loops
are shown in the next figure. The continuous line is the main feature of the diagram, separating
regions in which an abrupt nucleation event occurs (to the left of this line) or not (to the right). 
In addition, the nucleation event produces complete magnetization reversal to the left of the dotted line,
but only partial magnetization reversal between this curve and the continuous line.
}
\end{figure}

\begin{figure}
\includegraphics[width=8.5cm,clip=true]{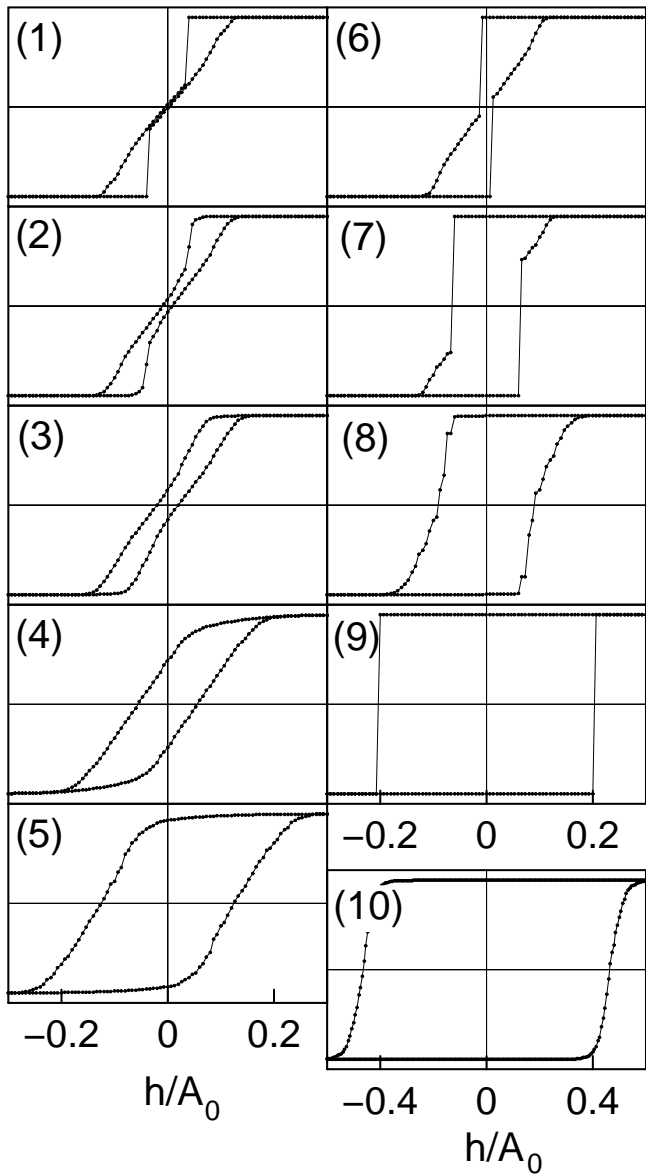}
\caption{\label{hyst}
The hysteresis loops at different locations of the phase diagram shown in the preceding figure.}
In the vertical axis we plot the magnetization $m_z$, which changes between $\pm 1$ as
the field is changed.
\end{figure}

The main result we are going to present is the disorder-exchange/anisotropy
ratio ($D$-$J/A_0$) phase 
diagram of the model, and its description. 
The phase diagram is presented in Fig. \ref{phasediag}. The form of the hysteresis loop 
at different positions is shown in Fig. \ref{hyst}.

There are three main qualitative regions in the phase diagram. 
For large $J/A_0$, ($2.1 \lesssim J/A_0$) the system shows no hysteresis at all.
A closer examination of the magnetization distribution shows that in this region the
interaction is so strong that effectively, the two minima of the local
magnetization disappear, and $m_z$ remains almost uniform in the sample, changing smoothly as
a function of
the applied field. This regime is certainly outside the region in which the model is justified, 
and then we will not consider it.
%This can be easily seen in the zero disorder case: assuming a uniform magnetization
%$\phi_0$, the gradient term is zero, and the dipolar term gives a quadratic contribution to the energy
%of the form $\phi_0^2\int_{\bf r}G({\bf r})$. When the coefficient of this term equals in absolute value the
%coefficient of the local energy term, the two wells of the  potential disappear, and the magnetization evolves
%continuously and non-hysteretically as a function of external field. This means that in this case,
%upon cycling of the field the system has a unique microscopic configuration
%at each value of the external field (note however that the existence of other--non accessed--metastable state
%is not ruled out).

More interesting to us is the rest of the phase diagram, where hysteresis occur, and where 
(from the examination of magnetization distribution) we observe well defined domains with 
magnetization $\pm 1$. That is the part plotted in Fig. \ref{hyst}.
A mayor distinction can be made within this region. It is related to the existence or not 
of a finite magnetization jump during field cycling.
The (numerically determined) regions 
in which a finite jump exists or not are separated
by the continuous line in Fig. \ref{phasediag}.
In addition, if a jump exists, it can lead directly to the fully inverted magnetization state,
or to a partial inversion, that requires a further change of the external field to completely
revert the magnetization. This two possibilities are separated by the dotted 
line in Fig. \ref{phasediag}.
A magnetization jump indicates an instability in the system, in which a finite 
fraction of the spins change their orientations, upon an infinitesimal change of the
external field. This avalanche process in magnetic systems was originally analyzed
in the random field Ising model (RFIM).\cite{sethna} It was shown
that if disorder in the system is lower than some critical amount,
then there is a magnetization jump, whereas there is no jump for higher disorder.
The same result is qualitatively obtained here, the amplitude of the magnetization jump vanishes
continuously when we approach the continuous line from the left in Fig. \ref{phasediag}.
This line is a critical line, in the same sense used in the RFIM. 
We expect that the vanishing of the abrupt jump follows a power law close to the critical line.
However, we have not attempted to determine critical exponents since in the presence of the dipolar interaction 
we cannot go to system sizes large enough to get good statistics. An interesting question that remains open
is whether the critical exponents are the same all along the critical line or not.

\begin{figure}
\includegraphics[width=8.5cm,clip=true]{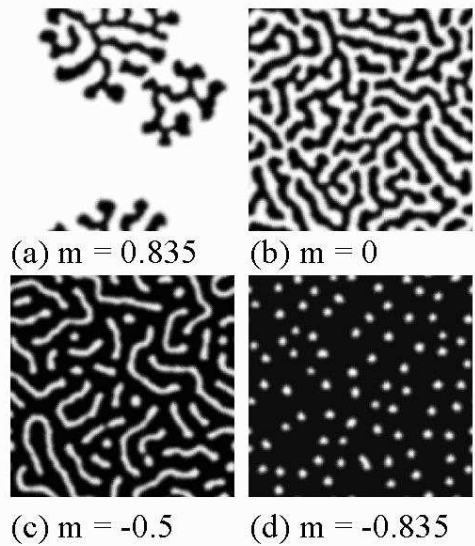}
\caption{\label{snaps1}Four snapshots of the spatial magnetization distribution corresponding 
to hysteresis loop (1) in Fig \ref{hyst}. The values of the magnetization at which the snapshot were
taken are indicated. Note that the first panel is an unstable configuration, which is spontaneously evolving
towards a state with $m\simeq 0.3$.}
\end{figure}

\begin{figure}
\includegraphics[width=8.5cm,clip=true]{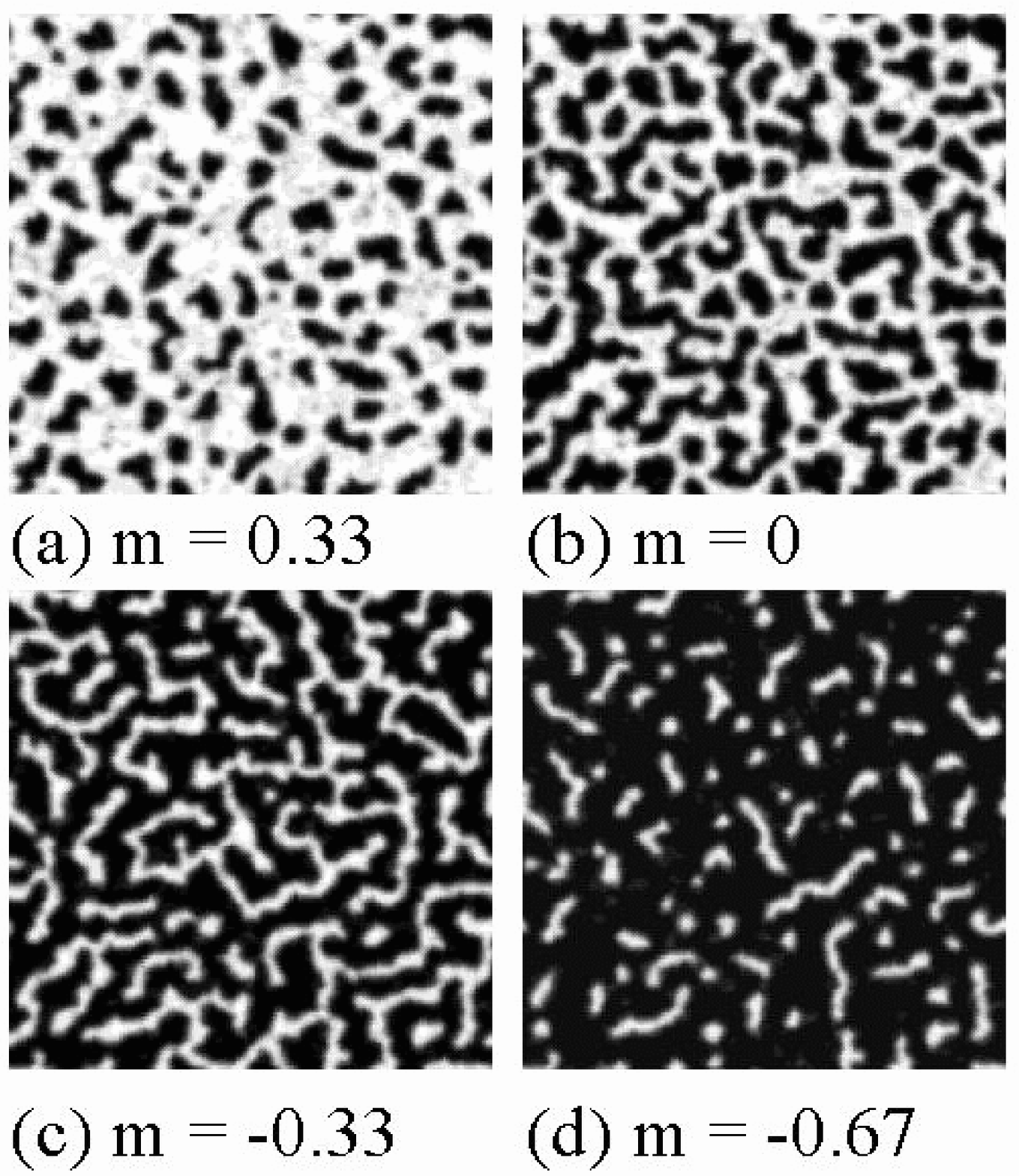}
\caption{\label{snaps2}Same as Fig. \ref{snaps1}, for hysteresis loop (4) in Fig \ref{hyst}.}
\end{figure}

\begin{figure}
\includegraphics[width=8.5cm,clip=true]{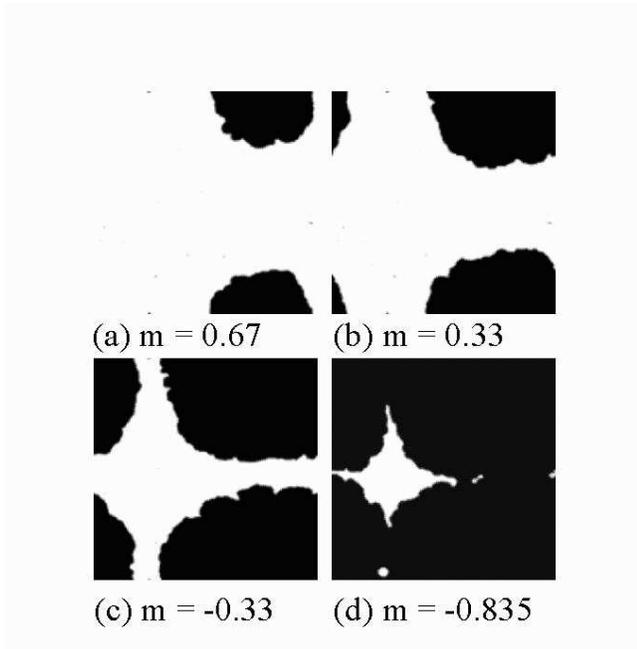}
\caption{\label{snaps3}Same as Fig. \ref{snaps1}, for hysteresis loop (9) in Fig \ref{hyst}. None of
these configurations is stable, they are spontaneously evolving to the completely inverted magnetization state.
}
\end{figure}

\begin{figure}
\includegraphics[width=8.5cm,clip=true]{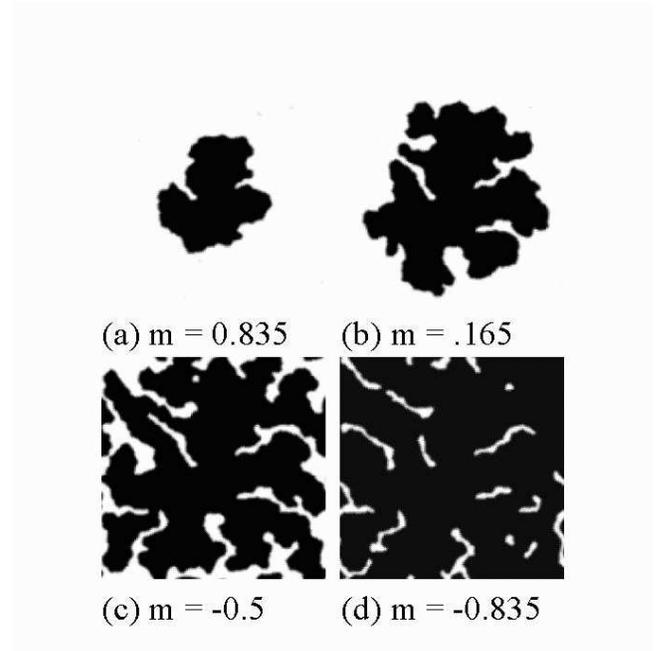}
\caption{\label{snaps4}Same as Fig. \ref{snaps1}, for hysteresis loop (7) in Fig \ref{hyst}.
First two panels show the spontaneously growing nucleated domain.}
\end{figure}

\begin{figure}
\includegraphics[width=8.5cm,clip=true]{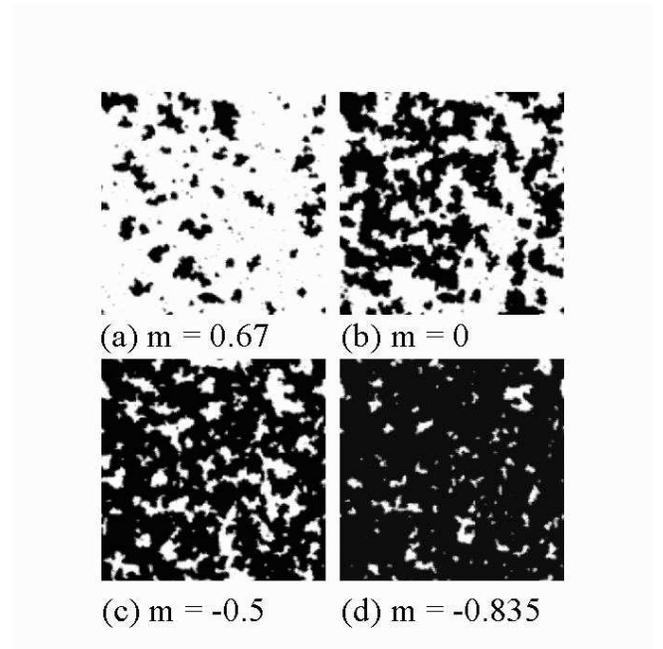}
\caption{\label{snaps5}Same as Fig. \ref{snaps1}, for hysteresis loop (10) in Fig \ref{hyst}.}
\end{figure}

The spatial distribution of the magnetization shows characteristic features
for different forms of the hysteresis loops.
Examples are shown in Figs. \ref{snaps1}, \ref{snaps2}, \ref{snaps3}, \ref{snaps4}, and \ref{snaps5}, 
corresponding to hysteresis loops (1), (4), (9), (7), and (10) in Fig \ref{hyst}.
Fig. \ref{snaps1} corresponds to a case where there is an abrupt nucleation, leading to a
partially inverted magnetization state. Panel (a) in Fig. \ref{snaps1} is a {\em non-equilibrium}
configuration after the nucleation. The final state at this field corresponds to
a configuration of magnetization $\sim 0.3$. Further decrease of the field (panels (c) and (d)) is necessary to completely invert the
magnetization. In the presence of strong disorder, for approximately the same value of $J/A_0$ (Fig. \ref{snaps2})
the nucleation step is smoothed, and the whole evolution is continuous. Here disorder plays a 
fundamental role
in preventing the spontaneous growth of the nucleated domains, which in this case remain pinned by the structural
disorder, and need a further decrease of the field to be able to grow.

Fig. \ref{snaps3} displays a case in which the nucleation events takes the sample completely in the oppositely
magnetized sated. Note how in this case, the domain that nucleates is a rather featureless, 
more or less circular objects, and grows to take over the whole sample. This case to be compared with
Fig. \ref{snaps1}, in which the domains nucleated maintain a striped internal structure.
An intermediate case is that of Fig. \ref{snaps4}. Here the inverted magnetization bubble almost invades the
whole sample at once. However, a small fraction of domain with the original orientation remains.
This is reflected in the form of the magnetization loop, which in fact shows an abrupt jump of the magnetization
from +1 to about -0.5.

The last case (Fig. \ref{snaps5}) corresponds to a much weakly interacting sample (lower value of $J/A_0$). However, the
disorder is sufficiently large to avoid a sudden nucleation step. We see a very different configurations of the
domains in this case, which do not show the typical striped pattern of the other cases (originated in the dipolar
interaction) but patterns mainly dominated by the disorder. This case, in which
the dipolar interaction play a minor role, is
qualitatively similar to that studied in the RFIM.

One striking characteristic of the phase diagram we are presenting
is the reentrance of the sector with continuous hysteresis. 
This reentrance tells that if $D$ is not too high,
samples displaying continuous hysteresis can be classified in two well different groups:
weakly interacting ones for low $J/A_0$, and strongly interacting ones for large $J/A_0$. 
For strongly interacting samples, the width of the hysteresis loop (measured as the distance between the two
branches, at a fixed magnetization) 
can be much smaller than the change of magnetic field necessary for complete 
magnetization inversion (see for instance
loop (3) in Fig. \ref{hyst}). In these 
samples the coercive field is much lower than the saturation field, 
and remmanent magnetization is much lower than saturation magnetization.

In weakly interacting samples the range of external field on which the complete magnetization inversion is
produced is small compared with the width of the hysteresis loop (as for instance in
loop (10) in Fig. \ref{hyst}). As a consequence, these samples
posses a coercive field which is of the order of the saturation field. In addition, the remmanent magnetization
is essentially the saturation magnetization.

\begin{figure}
\includegraphics[width=6.0cm,clip=true]{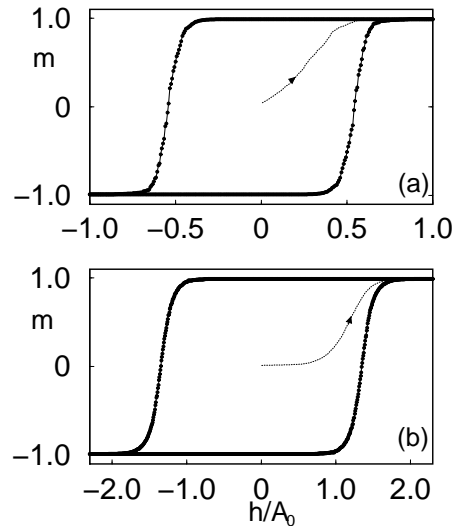}
\caption{\label{virgin}Hysteresis loop and virgin curve, obtained from a demagnetized sate, showing two different
qualitative cases: a `pinning dominated' magnet (a) in which the initial permeability of the sample is small, and 
a `nucleation dominated' magnet (b), with a large initial permeability. Note that both cases correspond to what
has been called "weakly interacting samples".}
\end{figure}

The case of what we have called `weakly interacting samples' allows yet a further distinction, which however
cannot be inferred from the results shown up to now only. In fact, weakly interacting samples are examples of
`permanent magnets'. A usual distinction among permanent magnets is that between nucleation type magnets an
pinning type magnets.\cite{huber} They are phenomenologically distinguished by the form of the virgin 
magnetization curve. Pinning type magnets display a small initial permeability at the demagnetized state,
whereas for nucleation type magnets the initial permeability is very large. The two different behaviors 
can be obtained with the present model. An example is shown in Fig. \ref{virgin}. 
(for simplicity in the simulation, we obtain a demagnetized system by {\em quenching} from a hight temperature
configuration, instead of the {\em annealing} process usually invoked).
Pinning type magnets are modeled
through a very low value of the interaction. The system behaves essentially as a collection of isolated magnetic
moments, that switch in the presence of the external field at the particular value of the local coercive field.
The transition is seen to be smoothed because of the disorder in the coercive field.
For larger values of the interaction strength (but still in the region corresponding to `weakly interacting
samples') we can see how the initial permeability increases noticeable. Here, domain walls (which are abundant in
the demagnetized state) have some freedom to move in the presence of the external field, and this produces a
high permeability. However, once fully magnetized and upon inversion of the external field, domains with opposite
magnetization have to nucleate, and this requires much larger fields. Note how the difference in the two virgin
curves occur in two samples that have an extremely similar appearance of the full hysteresis loop.

\section{Comparison with experiments and conclusions}

There has been in recent years much interest in the behavior of thin multilayer
magnetic films,\cite{exp1,exp2,exp3}
mainly as potential high density magnetic storage media.
Particular attention is being paid to Co-Pt multilayers.
The influence of disorder on these material has only recently started to be studied systematically.
An example of the evolution of hysteresis loops upon change of the structural disordered
in the sample is contained in [12]. %\cite{exp3}. 
There, Co-Pt multilayers were grown
under different values of argon sputtering pressure, what allows for the introduction of disorder
in a controlled way.
We can compare the plots in Fig. 1 of Ref. [12] %\cite{exp3} 
with the hysteresis loops for the
present model for a value $J/A_0\simeq 2$ (see panels (1) to (5) in Fig. \ref{hyst}) in which the 
disordered is progressively increased.
The qualitative agreement between the model and the experiment is astonishingly good,
taking into account the simplifications in the model. Comparison of our results  
with images from magnetic scattering techniques (Fig. 2 in Ref. [12], %\cite{exp3}, 
for instance) 
reveals also that even the real space configurations are quite realistic. 
We see that the samples analyzed in [11] and [12] %\cite{exp2} and \cite{exp3} 
fit very well in the strong interacting
region of our phase diagram.

The relevance of the present results to experiments is twofold. On one side, it provides insight on what are
the main physical ingredients that have to be considered to obtain a thorough description of the phenomenon. 
On the other side, it it also of importance in the following sense.
In many cases, experimental information of the real space patterns is
inferred from observation of the X-ray diffraction patterns, in which only the amplitude information
is conserved. Then the conclusions drawn from the experiments depend on the possibility of extracting
information of the real space patterns from the X-ray patterns, and this is not a trivial issue. The
present model, providing directly the configuration in real space (which can of course be transformed to get the
X-ray patterns) is an ideal benchmark in which the X-ray reconstruction techniques can be tested.
A more detailed statistical
comparison between simulated and measured patterns along these lines 
will be published elsewhere.\cite{pierce_unp}

\acknowledgments
I am indebted to Larry Sorensen and his group (at University of Washington), for persuading me 
to do simulations including disorder, and
for many exciting discussions and comments.

\end{document}